\documentclass[fleqn,usenatbib]{mnras}

\usepackage{newtxtext,newtxmath}

\usepackage[T1]{fontenc}
\usepackage{ae,aecompl}

\usepackage{graphicx}	
\usepackage{amsmath}	
\usepackage{amssymb}	
\usepackage{color}

\title[The integrated properties of the CALIFA galaxies]{The integrated properties of the CALIFA galaxies: \\ Model-derived galaxy parameters and quenching of star formation}

\author[T. Bitsakis et al.]{T. Bitsakis$^{1}$\thanks{CONACYT Research Fellow}\thanks{E-mail: t.bitsakis@irya.unam.mx},
S. F. S\'anchez$^{2}$,
 L. Ciesla$^{3,4}$,
P. Bonfini$^{5,6}$,
V. Charmandaris$^{5,6,7}$, \newauthor
B. Cervantes Sodi$^{1}$,
A. Maragkoudakis$^{8}$,
T. Diaz-Santos$^{9}$,
A. Zezas$^{5,6}$
\\
$^{1}$ Instituto de Radioastronom\'ia y Astrof\'isica, Universidad Nacional Aut\'onoma de M\'exico, Morelia, Mexico  \\
$^{2}$ Instituto de Astronom\'ia, Universidad Nacional Aut\'onoma de M\'exico, Mexico City, Mexico  \\
$^{3}$ Aix Marseille Univ, CNRS, CNES, LAM, Marseille, France \\
$^{4}$ Laboratoire AIM-Paris-Saclay, CEA/DSM/Irfu, CNRS, Universit\'e Paris Diderot, CEA-Saclay, 91191 Gif-sur-Yvette, France \\
$^{5}$ University of Crete, Department of Physics, 71003 Heraklion, Greece \\
$^{6}$ IESL/Foundation for Research and Technology-Hellas, 71110 Heraklion, Greece\\
$^{7}$ National Observatory of Athens, IAASARS, I. Metaxa and V. Pavlou, GR-15236, Penteli, Greece \\
$^{8}$ Department of Physics and Astronomy, University of Western Ontario, London, ON, N6A 3K7, Canada \\
$^{9}$ N\'ucleo de Astronom\'ia de la Facultad de Ingenier\'ia y Ciencias, Universidad Diego Portales, Av. Ej\'ercito Libertador 441, Santiago, Chile \\
}

\date{Accepted XXX. Received YYY; in original form ZZZ}

\pubyear{2018}

\begin{document}
\label{firstpage}
\pagerange{\pageref{firstpage}--\pageref{lastpage}}
\maketitle

\begin{abstract}
We present a study of the integrated properties of the 835 galaxies in the CALIFA survey. To derive the main physical parameters of the galaxies we have fitted their UV-to-IR spectral energy distributions (SED) with sets of theoretical models using {\sevensize CIGALE}. We perform a comparison of the integrated galaxy parameters derived from multi-band SED fitting with those obtained from modelling the Integral Field Unit (IFU) spectra and show the clear advantage of using the SED-derived star formation rates (SFR). A detailed analysis of galaxies in the SFR/M$_{star}$ plane as a function of their properties reveals that quenching of star formation is caused by a combination of gas deficiency and the inefficiency of the existing gas to form new stars. Exploring the plausible mechanisms that could produce this effect, we find a strong correlation with galaxy morphology and the build-up of central bulge. On the other hand, the presence of AGN and/or a stellar bar, as well as the local environment have only temporal effects on the current star formation, a result also consistent with their model-derived star formation histories.
\end{abstract}





\section{Introduction}
A long standing open problem of modern extragalactic astronomy concerns the mechanism(s) responsible for the morphological transformation and the quenching of star formation of spiral, gas-rich, galaxies into bulge dominated, gas-poor systems. Various quenching mechanisms have been proposed, such as:  $(i)$ the reduction of the gas content via stripping \citep{Quillis00} or strangulation \citep{Peng15}, $(ii)$ changes in the internal dynamics \citep[e.g. morphological evolution;][]{Martig09} and the enhancement of the central bulge \citep{Morishita15}, $(iii)$ feedback from an Active Galactic Nucleus \citep[AGN;][]{Fabian12}, or $(iv$) galaxy interactions and the local environment \citep[e.g. see][and references therein]{Struck99}. However, it is still unclear whether such processes may stimulate or suppress the star formation activity upon their action. For instance it is known that AGN outflows suppress star formation via the removal and/or the shock heating of the interstellar gas in their hosts \citep{Fabian12,Alatalo15}. However, it has been suggested that they may also enhance star formation via compression and subsequent fragmentation of the gas clouds \citep[e.g.][]{Barrera-Ballesteros15,Ellison18}. Various authors have also concluded that galaxy interactions (and collisions) enhance star formation activity in galaxy pairs \citep[e.g.][]{Smith07,Zaragoza18}, yet spiral galaxies in groups and clusters display significantly lower star formation compared to those found in the field \citep[see][]{Bitsakis11,Bitsakis16}. In addition, a strong correlation has been found between the total mass of the quenched galaxies and the epoch this process started \citep[i.e. high mass galaxies have been quenched at higher redshifts;][]{McDermid15,Pacifici16}.

Despite all those attempts to understand the star formation suppressing mechanisms, it is still unclear which of them dominate under the different circumstances and -- most importantly -- which are eventually responsible for quenching\footnote{Here with the term ``quenching'' we are referring to the permanent cease of star formation, independently if it was a rapid or a slow process.} the star formation. An important step towards the understanding of quenching is to study a statistically robust sample of galaxies that covers a wide range of masses, morphologies, and gas contents. It is also crucial to use a combination of state-of-the-art spectral energy distribution fitting which allows us to accurately constrain the integrated galaxy parameters (especially the star formation rates), with IFU spectroscopy that provides us with a wealth of spatially resolved data, essential to disentangle basic properties of the stellar populations and the gas. The prototypical such survey was the {\it Calar-Alto Legacy Integral Field spectroscopy Area} \citep[CALIFA;][]{sanchez12a}. This is a survey of $\sim$850 nearby galaxies using the state-of-the-art IFU mounted on the Calar-Alto 3.5m telescope. It is one of the largest, most complete IFU surveys \citep{walcher14} to study various aspects of galaxy evolution; there is also a wealth of multi-band photometric data for those galaxies. CALIFA still presents the best compromise between spatial coverage and resolution among the different IFU surveys. The targets are all nearby galaxies (z$<$0.03), covering a wide range of galaxy types and morphologies. 

In addition to its original aim to study star formation quenching this work also provides to the astronomical community the complete photometry as well as the integrated properties of all the galaxies, derived from theoretical modeling and IFU spectroscopy. This article is organized as follows: in \S2 we describe our sample and the data acquisition, in \S3 we present both methods used to obtain the integrated properties of the galaxies, i.e. the fitting of their observed spectral energy distributions with CIGALE as well as the estimated spectroscopic properties from the CALIFA pipeline. Our results are shown in \S4, while in \S5 we discuss the implications of those results and the overall picture we have on star formation quenching. Finally in \S6 we summarize our main conclusions. 

Throughout this work, a flat $\Lambda$CDM cosmological model is assumed, with parameters: H$_{0}$=70 km s$^{-1}$ Mpc$^{-1}$, $\Omega_{m}$=0.30, and $\Omega_{\Lambda}$=0.70.

\section{Sample}

The sample we use here comprises galaxies of the CALIFA survey. This is a survey of -- originally -- 630 (later extended to 1062) nearby galaxies observed with the IFU instrument of the Calar-Alto 3.5m telescope. The main criteria for the target selection of the original sample are fully described in \citet{walcher14}. The most important are:

\begin{itemize}
\item covered by the SDSS photometric survey;
\item angular isophotal diameter $45\arcsec - 79.2\arcsec$; 
\item redshift range $0.005<z<0.03$; 
\item Galactic latitude $|b| > 20\degr$; 
\item SDSS Petrosian magnitude r$< 20$; 
\item declination $\delta > 7\degr$. 
\end{itemize}

The lower redshift limit was imposed so that the original sample would not be dominated by dwarf galaxies. Therefore, CALIFA has a natural completeness limit at low mass. The upper redshift limit was imposed to guarantee that the spectral features necessary to define the stellar populations of the host galaxy were observable with the selected instrumental setup. The total volume probed by the CALIFA sample is $\approx10^6$ Mpc$^3$. Objects that are rare in number based on these selection criteria were targeted during one of the CALIFA ancillary programs, producing the extended sample \citep[consisting of 432 galaxies;][]{sanchez16c,Galbany18}. In that case some of the selection criteria were relaxed to allow the selection of particular targets such as dwarf galaxies, very massive ellipticals, supernovae hosts, merging galaxy companions etc.  

We acquire the largest possible multi-wavelength photometric dataset of all the objects within our sample from different public archives. By construction all  galaxies were included in the Sloan Digital Sky Survey-Data Release 7 \citep[{\it SDSS-DR7};][]{abazajian09}. Therefore, optical photometry is available for all of them in the u', g', r', i', and z' bands. The resolution is FWHM$\sim$1.4$\arcsec$ with sensitivities ranging between 22.0-20.5 mag for 5$\sigma$ detection. In this work we are using the photometry that was performed automatically by the SDSS pipeline using a combination of pure de Vaucouleurs and pure exponential profiles (SDSS model magnitudes). Near-infrared archival photometry was acquired from the Two-Micron All Sky Survey Extended Source Catalog archive \citep[{\it 2MASS-XSC};][]{skrutskie06}. The galaxies were imaged in the $J$, $H$, and $Ks$ bands, having 3$\sigma$ sensitivities of 14.7, 13.9, and 13.1 mag, respectively. The photometry was performed automatically by 2MASS using ellipsoidal apertures applied on the 3$\sigma$ J-band isophote of each galaxy. In addition, we obtain far-infrared archival data for a sub-sample of 501 galaxies from the AKARI Far-Infrared Survey \citep[{\it AKARI FIS};][]{okada08}, at the N60, WIDE-S, WIDE-L, and N160 filters, centered at 65, 90, and 140, and 160 $\micron$, respectively. The AKARI spatial resolution were FWHM$\sim$45$\arcsec$ at 65 and 90 $\micron$ and $\sim$60$\arcsec$ at 140 and 160 $\micron$ bands, respectively. 

Ultraviolet observations where obtained by the Galaxy Evolution Explorer \citep[{\it GALEX};][]{morrissey05} in the FUV (1540\AA) and NUV (2300\AA) bands simultaneously, with a spatial resolution of FWHM$\sim$5.5$\arcsec$. The photometry was performed by \citet{catalan15}. To calculate the integrated flux of each galaxy, these authors identified and removed foreground stars and other targets in the same field, replacing the masked areas by an interpolation along rows and columns from the adjacent non-masked pixels.

Finally, after a careful examination we decided against using the available archival data of the Wide Field Infrared Survey Explorer \citep[{\it WISE};][]{wright10} since they were originally optimized for point sources. We derived our own photometry using fixed apertures which where defined by \citet{catalan15} for the 22$\mu$m (i.e. the lowest-resolution) band of {\it WISE}. The final photometry used in the current work is presented in Table~\ref{tab1}. 

\begin{table*}
\begin{minipage}{120mm}
\begin{center}
\caption{The integrated photometry of the $CALIFA$ galaxies}
\label{tab1}
\hspace*{-4.0cm}
\begin{tabular}{ccccccccc}
\hline 
 CALIFA ID & Galaxy & redshift & GALEX FUV & GALEX NUV & SDSS u & SDSS g & SDSS r & SDSS i \\
           &  morphology &         &  [mag]    & [mag]     &  [mag]   &  [mag]    & [mag]   &  [mag]  \\
\hline
CALIFA001 & Sb & 0.016 & 17.86$\pm$0.15 & 17.23$\pm$0.05 & 15.57$\pm$0.10 & 13.92$\pm$0.05 & 13.20$\pm$0.05 & 12.76$\pm$0.05 \\      
CALIFA002 & Sbc & 0.023 & 16.67$\pm$0.01 & 16.18$\pm$0.04 & 14.94$\pm$0.10 & 13.58$\pm$0.05 & 12.97$\pm$0.05 & 12.63$\pm$0.05  \\
... & \\
\hline
\end{tabular}

\vskip 0.2cm

\hspace*{-4.0cm}
\begin{tabular}{ccccccccc}
\hline 
   & SDSS z &  2MASS J & 2MASS H & 2MASS Ks & WISE 3.6 & WISE 4.6 & WISE 12 & WISE 22 \\
   & mag] & [mag] & [mag] & [mag] & [mag] & [mag] & [mag] & [mag] \\
\hline
   & 12.54$\pm$0.10  & 12.29$\pm$0.02 & 12.09$\pm$0.02 & 12.25$\pm$0.03 & 12.59$\pm$0.01 & 13.12$\pm$0.01 & 12.06$\pm$0.01 & 12.49$\pm$0.06 \\      
   & 12.41$\pm$0.10 & 12.03$\pm$0.02 & 11.90$\pm$0.03 & 12.04$\pm$0.04 & 12.40$\pm$0.01 & 12.86$\pm$0.01 & 11.03$\pm$0.01 & 10.64$\pm$0.02 \\
   & .. & \\
\hline
\end{tabular}
\end{center}
\end{minipage}
\end{table*}%
\section{Estimation of the physical properties of the galaxies}

\begin{table*}
\begin{minipage}{120mm}
\begin{center}
\caption{The integrated galaxy properties}
\label{tab2}
\begin{tabular}{cccccc}
\hline 
 CALIFA ID & log M$_{star}$ & SFR &  log M$_{dust}^{a}$  &  log M$_{H2}$ & f$_{AGN}^{b}$\\
        & [M$_{\odot}$]  & [M$_{\odot}$ yr$^{-1}$] & [M$_{\odot}$]  & [M$_{\odot}$] & \\
\hline
CALIFA001 & 10.75$\pm$0.06 & 0.51$\pm$0.11 & - & 9.23$\pm$0.05 & - \\
CALIFA002 & 11.10$\pm$0.06 & 4.76$\pm$1.21 & 8.70$\pm$0.56 & 9.93$\pm$0.20 & - \\
... & \\
\hline
\end{tabular}
\end{center}
NOTES: $^{a}$ Available only for galaxies with FIR observations.\\ $^{b}$ The fraction of AGN-to-total IR luminosity.
\end{minipage}
\end{table*}%

\subsection{Spectral Energy Distribution fitting}
We used the Code for Investigating Galaxy Emission ({\sevensize CIGALE}; \citealt{Noll09}) to estimate the physical properties of the 1062 galaxies in the CALIFA survey. The code uses a Bayesian-like approach to fit the observed SEDs of the galaxies with sets of theoretical models, taking into account the balance between the energy absorbed in the UV and re-emitted in the infrared. It then builds the probability distribution functions of the physical parameters of the galaxies, such as the stellar mass, the star formation rate, and the extinction. The assumption on the star formation history is versatile. Here we use a delayed-$\tau$ model but allowing for large variations ($\Delta$log(SFR)=10) in the recent star-formation history (SFH), to account for the most extreme changes such as a starburst, a rejuvenation of star formation, or fast quenching, as well as moderate variations within the star formation main sequence. This SFH is intensively tested and described in more detail in \cite{ciesla16} and \cite{ciesla17} where they show that it provides accurate SFR (averaged over 10 \& 100 Myr), especially in the case of starburst or recently quenched galaxies.  {\sevensize CIGALE} also models the presence of active galactic nuclei (AGN), using three different types of nuclear activity (from type 1, type 2 and intermediate type AGN) to fit -- if present -- the non-SF nuclear emission.  However, since we are using broad-band photometry the AGN models might be degenerate. \citet{Ciesla15} reported that these models can be reliable only for fractions of AGN-to-total IR luminosity (f$_{AGN}$) of $>$10 per cent. Moreover, the models become degenerate when using both AGN contribution and a flexible SFH. Therefore, we fit all galaxies with either $(i)$ a flexible SFH or $(ii)$ a fixed SFH with AGN contribution, and then select the best model in each case based on their $\chi^{2}$. We have to stress here that the variation in the galaxy parameters (especially those related to the infrared luminosity, such as the SFR) between the two models is always within the uncertainties for all galaxies without a significant contribution of an AGN in their infrared colours. In Table~\ref{tab2} we present the parameters of the 1062 galaxies. Dust masses are only estimated for those galaxies with available FIR observations.

\subsection{The CALIFA dataproducts}
Along this study we use the results of the analysis of the CALIFA datacubes (covering the wavelength range 3700 $\leq\lambda<$ 7000 \AA) provided by the {\sevensize Pipe3D} pipeline \citep{sanchez16b}. This pipeline automatically provides the stellar masses, star formation rates (SFRs), dust attenuation (for both the ionized gas and the stellar population), and the star-formation histories of galaxies. Here we present a brief summary of the more relevant steps performed in the context of our current analysis.

Initially, a binning procedure is adopted to increase the signal-to-noise ratio of the spaxels up to a certain required threshold ($>$50). This S/N limit was selected based on extensive simulations in order to recover reliably the SFHs of the different regions in the galaxies \citep{sanchez16b}. Although a substantial number of spaxels have already S/N larger than this limit, for the remaining ones it is necessary to increase their S/N above the required limit by co-adding the adjacent spectra into a spatial bin (a.k.a. ``tessellas''; typically 2-5 spaxels in size). The adopted binning scheme has the advantage of providing  an inhomogeneous S/N distribution which better preserves the original borders of the galaxy; all galaxies are sampled to $>$2 effective radii (R$_{eff}$) with 90 percent up to 2.5 R$_{eff}$. Moreover, although the tessellas with the lowest S/N typically correspond to the outer regions of the galaxies, the radially averaged or integrated properties across the entire FoV have similar uncertainties to those of the spaxels with the highest S/N \citep[for more details see][]{ibarra16}. \citet{Sanchez16a} show how successfully are recovered those SFHs by the pipeline; for a single burst in Fig. 7, for a typical SFH and metal enrichment history in Fig. 9, and for a realistic SFH and metal enrichment history in Fig. 15. Once performed the spatial binning/segmentation, the spectra from the spaxels of each tessella are co-added prior to any further analysis.

The fitting procedure includes both, stellar continuum and emission line modelling, in an iterative processes where the former is fitted first, masking out the emission lines. Then, the emission lines are modelled using single Gaussian functions. This iterative process is repeated until the chi-square value is minimized. To derive the stellar mass, {\sevensize Pipe3D} fits the underlying stellar population of each binned spectrum with a library of single-stellar populations \citep[SSPs;][]{cid-fernandes13} that comprises 156 templates, covering 39 stellar ages (from 1Myr to 14.1Gyr) and 4 metallicities (Z/Z$\odot$=0.2, 0.4, 1, and 1.5). The contribution of each SSP of a certain age and metallicity is derived by a linear inversion procedure for each of the  spectra. Then, the mass distribution is derived along those components by taking into account the M/L ratio of each SSP within the library. By co-adding the metallicity range we obtain the amount of stellar mass distributed along the different considered ages. Each age corresponds to a certain look-back time, taking into account the redshift of the galaxy. From this distribution, it is possible to derive the spatially resolved cumulative mass function along the cosmic look-back time, integrating sequentially along the ages, and the spatially resolved SFH, as the derivative of this cumulative function, once corrected for the mass loss. 
We also estimate the current SFR as the amount of stellar mass formed in a certain recent time bin divided by the length of that bin, following \citet{rosa16a} and \citet{sanchez18}.

In addition, the stellar population model is used to subtract the stellar continuum in each spaxel and provide with a gas-pure datacube. From this cube we estimate the properties of a set of emission line species, including the more relevant to: $(i)$ determine the nature of the ionization, $(ii)$ derive the dust attenuation in the gas, and $(iii)$ provide the H$\alpha$-derived SFR. 
Global properties, like the integrated SFR or the average dust attenuation are derived by either co-adding or averaging the corresponding spatial resolved values within the FoV of the datacubes. Eventually, 937 galaxies were observed within the timeframe of the CALIFA survey and from those, 835 where successfully fitted by {\sevensize Pipe3D}. 

Molecular hydrogen masses of 126 (109 of which are late-type) galaxies in our sample were derived from the CO(1-0) observations of the Extragalactic Database for Galaxy Evolution survey \citep[EDGE;][]{Bolatto17}. These galaxies were observed using the Combined Array for Millimeter-wave Astronomy (CARMA), with a spatial resolution of $\sim$1.4 kpc for all galaxies. For the remaining 709 galaxies the molecular gas masses were inferred from the CALIFA pipeline using a relation between the derived Av and the H$_{2}$ column densities \citep[see][and references therein]{Galbany17}. A comparison between the Av and the CO derived ones for the 126 galaxies shows an excellent agreement between the two methods \citep{Barrera-Ballesteros18}, having a very small scatter of 0.08 dex that does not depend on galaxy morphology or metallicity. 

We also use the emission line diagnostic diagrams to classify the nuclear activity of the CALIFA galaxies. 
We apply our analysis at a central spaxel of each galaxy (3"$\times$3"), following \citet{sanchez18}. The AGN selection criteria we use are the following: $(i)$ the H$\alpha$ equivalent width (EW(H$\alpha$)) $>$6\AA,  to ensure that the emission lines originate from a strong excitation field rather than post-AGB stars (typically having EW(H$\alpha$)$\le$2\AA~\citep[e.g. see][]{Binette94,stas08}. $(ii)$ The emission line ratios of the galaxies must be located above the \citet{kewley06} demarcation lines, in all three ([O{\sevensize III}]5007/H$\beta$ versus the [N{\sevensize II}]6583/H$\alpha$, or the [S{\sevensize II}](6717+6731)/H$\alpha$, or the [O{\sevensize I}]6300/H$\alpha$) emission line diagrams. 
Following the above criteria only 18 galaxies are classified as AGN hosts. We should note here that our selection criteria are quite strict, and relaxing them could increase considerably the number of AGNs, as has been found in other sample of galaxies \citep[e.g.][]{sanchez18}. However, with these criteria we ensure that our selected sample comprises bona-fide optically selected AGNs.

In Table~\ref{tabsample} we present a summary of all those samples mentioned above.


\begin{table*}
\begin{minipage}{120mm}
\begin{center}
\caption{Sample summary}
\label{tabsample}
\begin{tabular}{lc}
\hline 
Description & No of galaxies \\
\hline
CALIFA mother sample & 630 \\
CALIFA extended sample & 432 \\
CALIFA observations that passed the data quality control  & 835 \\
Galaxies modelled by Pipe3D  & 835\\
Galaxies modelled by CIGALE & 1062\\
Galaxies with CO-derived H$_{2}$ masses & 126 \\
Galaxies with model-derived H$_{2}$ masses & 709 \\
AGN hosts & 18 \\
\hline
Final sample considered in this work & 835 \\
\hline
\end{tabular}
\end{center}
\end{minipage}
\end{table*}%

\begin{figure}
\begin{center}
\includegraphics[scale=0.47]{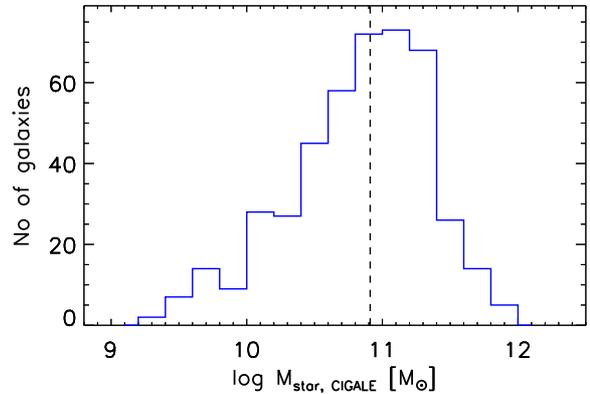}
\caption{The distribution of stellar masses of the galaxies in the CALIFA survey. The dashed vertical line indicates the median stellar mass at log(M$_{star}/$M$\odot$)=10.9. (A coloured version of this figure is available in the online journal)}
\label{fig_mass_distr}
\end{center}
\end{figure}

\begin{figure}
\begin{center}
\includegraphics[scale=0.47]{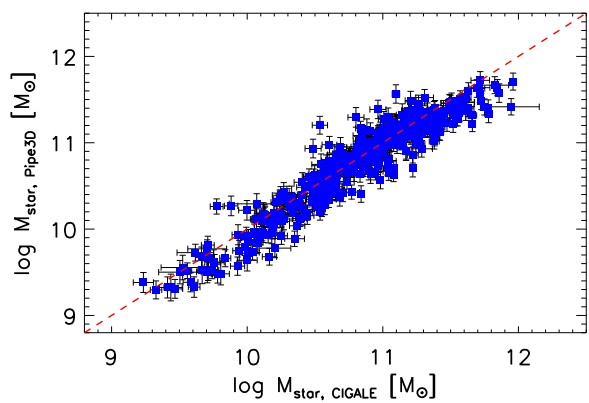}
\caption{Comparison of the SED-derived stellar masses derived from {\sevensize Pipe3D} and {\sevensize CIGALE}. The dashed line marks the one-to-one correlation. (A coloured version of this figure is available in the online journal)}
\label{fig_mstar}
\end{center}
\end{figure}

\section{Results}

\subsection{Presenting the integrated galaxy properties}

\begin{figure}
\begin{center}
\includegraphics[scale=0.47]{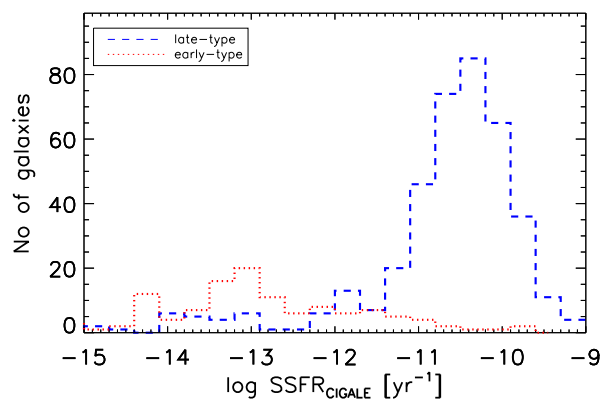}
\caption{The distributions of the SED-derived specific star formation rates of the late-type (dashed blue) and early-type (dotted red) galaxies in our sample. (A coloured version of this figure is available in the online journal)}
\label{fig_SSFR_distr}
\end{center}
\end{figure}

\begin{figure}
\begin{center}
\includegraphics[scale=0.47]{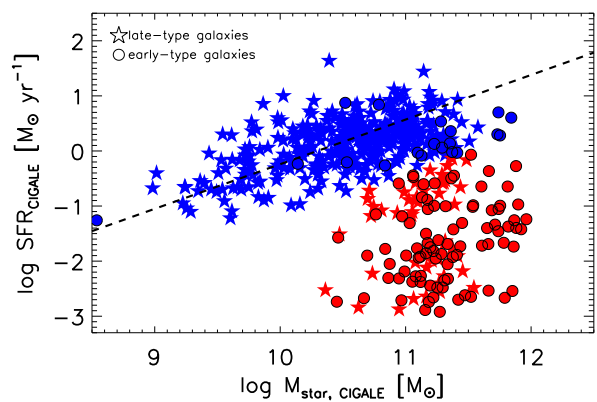}
\caption{The SFRs of the CALIFA galaxies estimated by SED modeling as a function of their stellar mass. Late-type galaxies are shown with stars and early-type galaxies with circles. Star forming galaxies are marked with blue symbols (log(sSFR/yr)>-11.5), whereas quiescent galaxies are marked with red. The dashed black line indicates the star forming main sequence derived from \citet{Elbaz07}. (A coloured version of this figure is available in the online journal)}
\label{fig_SFMS}
\end{center}
\end{figure}

In this section we present the integrated galaxy properties estimated from the SED fitting and compare them with those derived from the CALIFA pipeline. In Fig.~\ref{fig_mass_distr}, we present the distribution of the {\sevensize CIGALE}-derived stellar masses of our galaxies, while in Fig.~\ref{fig_mstar} we compare them with those from {\sevensize Pipe3D}. Albeit the very good agreement between the two estimators (having a scatter of 0.05 dex), there is an small offset of 0.04 dex. Since estimators are using a Salpeter IMF, this discrepancy may arise from the field-of-view limitations of the IFU instrument, since CALIFA is systematically missing the flux coming from the more distances outskirts of the galaxies. This results in the under-estimation of the pipeline-derived stellar masses, and suggests that even covering beyond 2.5 R$_{eff}$ for most of the galaxies \citep[see][]{walcher14}, we still detect a small aperture effect.

Similarly, in Fig.~\ref{fig_SSFR_distr} we present the distributions of the specific star formation rates (the star formation rate over the stellar mass of a galaxy; sSFR) of the late-type (dashed blue) and early-type (dotted red) galaxies in our sample. We can see that there are several -- morphologically classified -- late-type galaxies which display very low sSFRs, comparable to those of early-type galaxies, while the SF galaxies are found in the range -11.5<log(sSFR/yr)<-9, according to the sSSFR distribution. To further study this peculiarity we plot in Fig.~\ref{fig_SFMS} the relation between the SFR and the stellar mass. We classify our galaxies as ``star-forming'' (SF) if they lie on the well-known star formation main-sequence \citep[SFMS;][]{Elbaz07}, and as ``quiescent'' if they are found below it (having log(sSFR/yr)<-11.5). We find that many morphologically classified late-type galaxies are in fact quiescent. This seems -- at first -- an odd result that requires a deeper analysis and interpretation (see \S4.2).

\begin{figure}
\begin{center}
\includegraphics[scale=0.47]{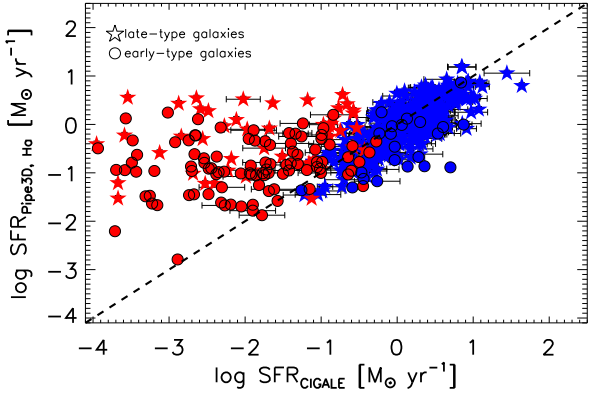}
\caption{Comparison of the H$\alpha$-derived star formation rates from {\sevensize Pipe3D} and those from {\sevensize CIGALE}. Late-type galaxies are shown with stars and early-type galaxies with circles. Star forming galaxies (as indicated in Fig.~\ref{fig_SFMS}) are marked with blue symbols, whereas quiescent galaxies are marked with red. The dashed black line marks the one-to-one correlation. (A coloured version of this figure is available in the online journal)}
\label{fig_SFR}
\end{center}
\end{figure}

In Fig.~\ref{fig_SFR} we compare the H$\alpha$-derived star formation rates from {\sevensize Pipe3D} with those from {\sevensize CIGALE}. Although, there is a relatively good correlation (dispersion 0.3 dex) for SFRs above 0.1 M$_{\odot}$ yr$^{-1}$, the relation does not hold for lower values. One can see that those outliers are always quiescent galaxies (according to Fig.~\ref{fig_SFMS}), suggesting that the significant divergence between the two SFR estimators should arise from the different assumptions they both make. In \S3.2, it is mentioned that {\sevensize Pipe3D} is using the H$\alpha$ fluxes to infer the recent ($<$10 Myr) SFRs of the galaxies. This presumes that the integrated H$\alpha$ emission of a given galaxy originates from the excitation of gas from young stars. However, this is only true for SF galaxies, whereas there are various alternative sources of excitation for quiescent galaxies \citep[e.g.][]{Singh13}. These include the emission from an active galactic nuclei (AGN), or evolved stars \citep[e.g.][]{Binette94,Sarzi10}; collisional shocks could also produce such an excitation \citep[e.g.][]{Alatalo16}. \cite{mariana16} studied the (spatially resolved) SFMS of the CALIFA galaxies and argued for the  existence of an additional ``main sequence'' of quiescent galaxies in their data. Similar results has been also observed \citep{Hsieh17} in the ``Mapping Galaxies with APO'' survey  \citep[{\it MaNGA};][]{Bundy15}. The presence of such a quiescent galaxy sequence disappears when we use the SED-derived SFRs (see our Fig.~\ref{fig_SFMS}). \cite{mariana16} have thus assumed that this main sequence of quiescent galaxies is probably due to the ionized gas surrounding evolved stars rather than gas excited by young stars, and this is why it scales linearly with the stellar mass, i.e. the more massive a galaxy, the more evolved stars it contains. On the other hand, such a linear scaling does not necessarily hold for AGN-hosting quiescent galaxies, since their H$\alpha$ emission depends on the activity stage of the AGN which does not necessarily scale with the galaxy mass. These assumptions also agree with the results of \citet{Kehrig12}, \citet{Papaderos13}, and \citet{Gomes16} who studied individual CALIFA early-type galaxies and proposed that evolved stars are the major sources of excitation.

\begin{figure}
\begin{center}
\includegraphics[scale=0.47]{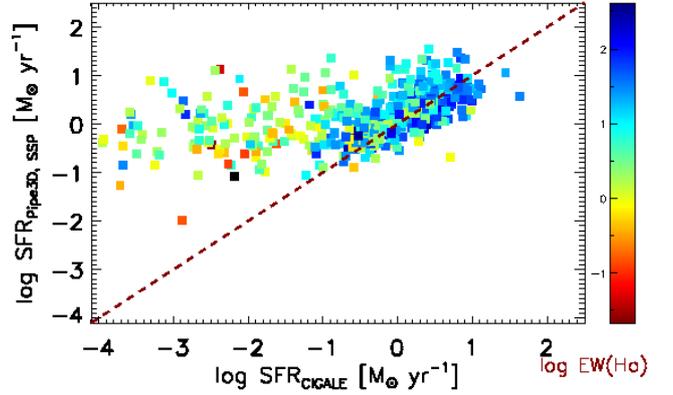}
\caption{Comparison of the SSP-derived star formation rates from {\sevensize Pipe3D} and those from {\sevensize CIGALE} as a function of their EW(H$\alpha$). The dashed black line marks the one-to-one correlation. (A coloured version of this figure is available in the online journal)}
\label{fig_SFR_ssp}
\end{center}
\end{figure}

Nevertheless, in \S3.2 it is also mentioned that {\sevensize Pipe3D} is able to estimate the SFRs of the galaxies by fitting single stellar population (SSP) models in the spectra of our galaxies. In Fig.~\ref{fig_SFR_ssp} we compare those with the SED-derived SFRs, with the galaxies color-coded as a function of their EW(H$\alpha$). Although this is a completely different SFR estimator, we can still see a large discrepancy. This suggests that not even optical spectroscopy is able to distinguish between post-AGB and young stars (at least at the wavelength range covered by CALIFA; 3700-7000 \AA) with result to over-predict the contribution of the latter. This is an interesting result that was also presented in \citet{LopezFernandez16}, who performed simultaneous photometric and spectroscopic analysis in a sample of 260 galaxies extracted from CALIFA, using the stellar synthesis code {\sevensize STARLIGHT}, and concluded that to correct for this overestimation, UV photometric data must be used in addition to the optical spectroscopy. Yet, as indicated in our figure, galaxies with low EW(H$\alpha$) -- typically lower than 3\AA~\citep[see][]{Lacerda18} -- are those displaying the largest discrepancies and, thus, one could use that information given by {\sevensize Pipe3D} to discard the outliers.


\begin{figure}
\begin{center}
\includegraphics[scale=0.47]{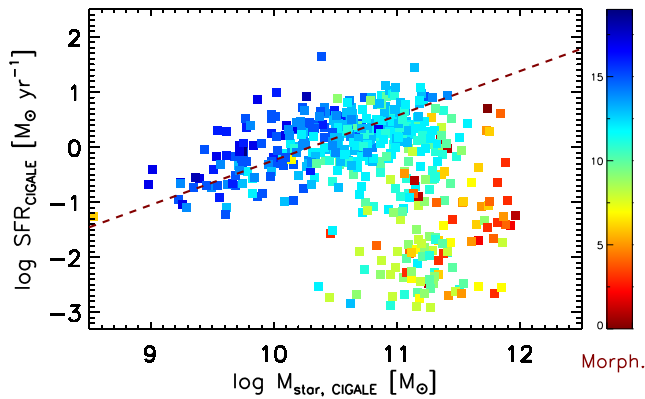}
\caption{The SFRs of the CALIFA galaxies estimated by SED modeling versus their stellar mass, as a function of the galaxy morphology. Elliptical galaxies have Morphological indices $\leq$7, lenticulars 8-9, and spirals $\ge$10 \citep[see][]{walcher14}. (A coloured version of this figure is available in the online journal)}
\label{fig_morph}
\end{center}
\end{figure}

\subsection{Disentangling the Star-Formation Main Sequence}
As mentioned in the previous paragraph, an interesting result derived from Fig.~\ref{fig_SFMS} is that there are plenty of -- morphologically classified -- late-type galaxies (LTGs) that are located below the SFMS. In order to investigate this irregularity, we plot in Fig.~\ref{fig_morph} the stellar mass versus SFR diagram of our galaxies as a function of the galaxy morphologies \citep[these are visual classifications performed by members of the CALIFA collaboration; see caption of Fig.~\ref{fig_morph} and][ for more details on the classification indices]{walcher14}. Two results are evident in that figure: {\it(i)} the evolutionary sequence of galaxy morphologies that varies from late-type spiral into earlier type spiral, lenticular, and finally elliptical galaxies, and {\it(ii)} the bimodality in the sequence of quiescent galaxies, where early-type spirals and lenticulars are mostly located at log(M$_{star}$/M$\odot$)$<$11.4, whereas ellipticals are more massive. 
However, it is not obvious how to track this evolutionary sequence. Is the build-up of stellar mass and the subsequent  morphological evolution (e.g. via the formation of a bulge) responsible for quenching the star formation activity in those galaxies \citep[e.g.][]{Schreiber16}, or should we also consider other processes -- internal or external -- responsible for this (e.g. AGN or gas stripping)?

\begin{figure}
\begin{center}
\includegraphics[scale=0.47]{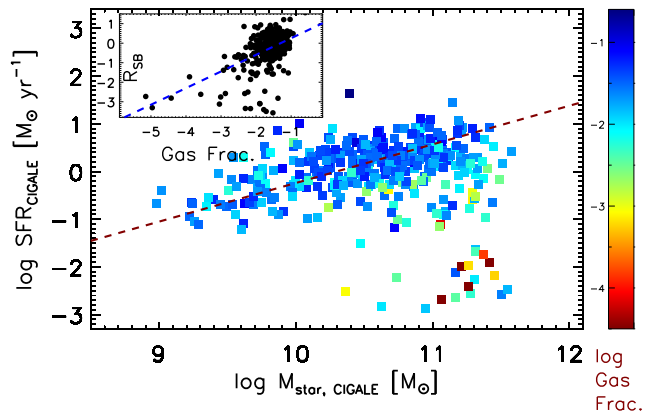}
\caption{The SFRs of the CALIFA late-type galaxies estimated by SED modeling versus their stellar mass, as a function of the gas fraction, i.e. the molecular gas mass of each galaxy normalized by its stellar mass, presented in log scale. In the inset panel we plot the R$_{SB}$ versus the gas fraction for the same galaxies. The dashed line indicates the linear fit of the data. (A coloured version of this figure is available in the online journal)}
\label{fig_gfrac}
\end{center}
\end{figure}

In Fig.~\ref{fig_gfrac}, we present the stellar mass versus the SFR diagram of the LTGs in our sample, colour-coded by their molecular gas mass fraction, which is the molecular gas mass of each galaxy normalized by its total mass (gas+stellar). This fraction can be considered as a proxy for evolutionary stage in that it compares the mass already converted into stars with the mass which is still in gaseous form. The gas masses of 106 of those galaxies were derived using the CO(1-0) observations of the Extragalactic Database Galaxy Evolution survey \citep[EDGE;][]{Bolatto17}. For the remaining 266 galaxies we have used the simple -- but accurate -- correlation between the A$_{v}$ and the H$_{2}$ column density as explained in \S3.2.  According to Fig.~\ref{fig_gfrac}, the LTGs below the SFMS are those having the lowest gas fractions. In the inset panel we present the distance from the SFMS, R$_{SB}$ \citep[which is given by the ratio of the SFR of each galaxy, over the SFR it would have if it was located on the SFMS;][]{Elbaz07}, as a function of the gas fraction. If gas deficiency is the only factor regulating star formation, those two values are expected to scale, as it is the case for all the SFMS galaxies (fitted by the dashed blue line). Yet for most LTGs located below the SFMS their R$_{SB}$ does not scale with the deficiency of gas, suggesting the presence of additional suppressing mechanisms. A similar conclusion has been reached by \citet{Saintonge16}, who studied the SFMS plane of a large sample of nearby SF galaxies and showed that the deficiency of atomic and/or molecular gas is not the only responsible factor for driving sSFR in those galaxies. They attributed the additional discrepancy to the decline of the star formation efficiency as well as in variations of the molecular-to-atomic gas mass ratio. 

\begin{figure}
\begin{center}
\includegraphics[scale=0.47]{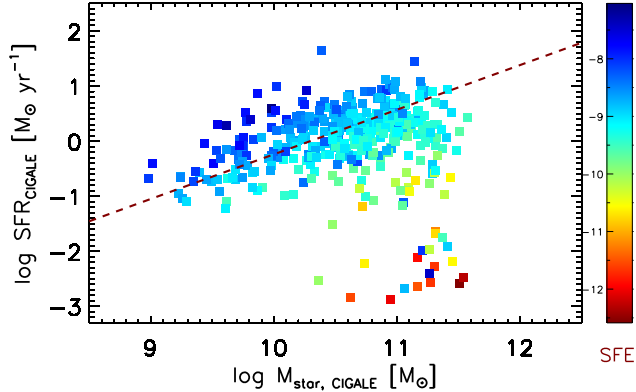}
\caption{The SFRs of the CALIFA late-type galaxies estimated by SED modeling versus their stellar mass, as a function of the star formation efficiency, which is the log(SFR/M$_{H2}$) [yr$^{-1}$]. (A coloured version of this figure is available in the online journal)}
\label{fig_eff}
\end{center}
\end{figure}

Indeed in Fig.~\ref{fig_eff} we plot the SFR/M$_{star}$ plane of the LTGs in our sample, this time colour-coded by their star formation efficiency (SFE). This is the  ratio of the SFR over the molecular gas mass, and it is a measure of the efficiency with which galaxies are using their remaining gas to produce stars (the inverse ratio is simply the depletion time, i.e. the time in which a galaxy will consume the  available amount of gas with the current rate of star formation). The plot confirms our previous assumption that LTGs above the SFMS are forming stars more efficiently than those below it. Moreover, the farther below a galaxy is from the SFMS, the more inefficient it is in forming stars. This relation holds even if we only consider the LTGs for which we have CO-derived gas masses. This result complements the work of \citet{Scoville17} for galaxies at z = 0, who showed that the SFE of galaxies above the SFMS increases as the 0.7 power of their R$_{SB}$. They also found a dependence between the gas fraction and the elevation of galaxies above the SFMS. Our findings in Fig.~\ref{fig_gfrac} indicate that this trend also extends to galaxies below it. However, a larger sample of galaxies in this region of the parameter space are needed to perform a similar analysis to \citet{Scoville17} and obtain the precise power-law dependencies on the parameters that control the position of galaxies in the SFR/M$_{star}$ plane.

Combining the above results we can conclude that galaxies are found below the SFMS because of at least one of the following factors: $(i)$ the deficiency of molecular gas, and/or $(ii)$ the inefficiency of the remaining gas to form stars.
Various mechanisms, both internal and environmental, have been proposed to have such effects on galaxies. AGN feedback \citep{Fabian12}, and morphological evolution \citep{Martig09} are some important internal mechanisms. On the other hand, galaxy interactions and the kinetic energy induced to the gas by collisional shocks \citep{Alatalo15,Bitsakis16} are some of the environmental ones. 

\subsection{Exploring the various processes that may influence star formation}

\begin{figure}
\begin{center}
\includegraphics[scale=0.47]{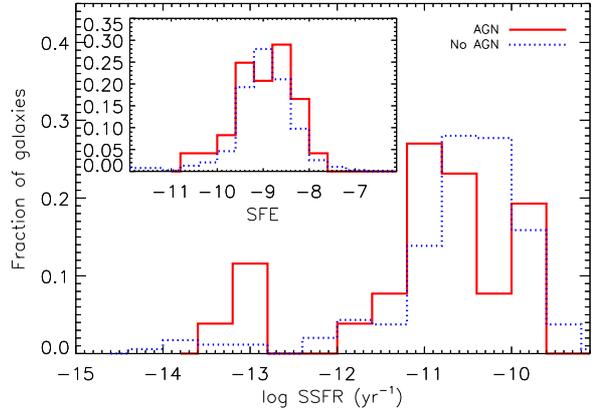}
\caption{The sSFR distributions of the late-type galaxies hosting an AGN according to both SED and emission-line classifiers (solid red) in comparison with those that they do not contain one (dotted blue). In the inset panel we present the respective distributions of star formation efficiency. (A coloured version of this figure is available in the online journal)}
\label{fig_AGN_distr}
\end{center}
\end{figure}

Although AGN is associated with suppressing the star formation activity in galaxies \citep[or at least in their central parts;][]{catalan17} their duty cycle is short and might not have a permanent influence on quenching. To explore this possibility, we plot in Fig.~\ref{fig_AGN_distr} the distributions of the sSFRs of the AGN-hosting LTGs in our sample against those without a nuclear source, classified by both SED and emission-line diagnostics (see \S3). A Kolmogorov-Smirnov (KS)-test reveals that the two distributions are different (P$_{KS}\sim$0.02); in fact we can see that the incidence of AGNs is higher among lower sSFR LTGs. There might be various explanations for that, such as: $(i)$ the AGN are responsible for suppressing the star formation activity \citep[e.g. see][and references therein]{Fabian12}, $(ii)$ AGN is a process simultaneous to quenching, or $(iii)$ weak AGN are more possible to be seen when star formation fades-out \citep[e.g.][]{Bitsakis15}. To resolve that, we plot in the inset panel of the same figure the star formation efficiency distributions of the AGN and non-AGN galaxies. The two distributions do not display any difference, suggesting AGN do not seem to be responsible for suppressing the integrated star formation efficiency of the gas in our galaxies \citep[in contrast to what is found in quasars;][]{dimatteo05}. Yet an alternative scenario would have been the AGN to produce a chain of effects that eventually will quench star formation (e.g. starvation due to feedback), but since AGN timescales are much shorter than the quenching process, it is difficult to ascribe a causal connection between the two. Although appealing, we do not have any evidence to support that scenario.

\begin{figure}
\begin{center}
\includegraphics[scale=0.47]{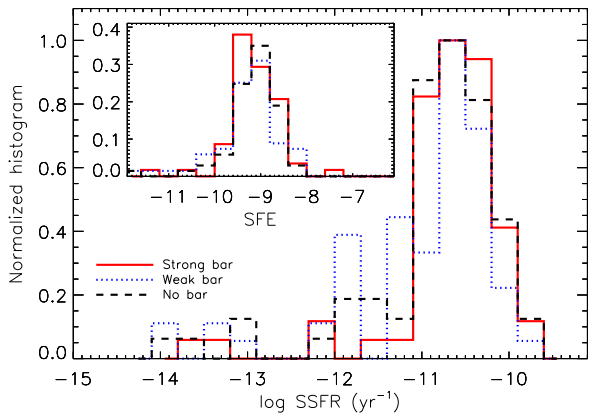}
\caption{Comparison of the sSFR distributions of the LTGs in our sample which display a prominent (solid red line) or a weak (dotted blue line) bar, against those without a bar (dashed black line). In the inset panel we present the respective distributions of star formation efficiency. (A coloured version of this figure is available in the online journal)}
\label{fig_bar}
\end{center}
\end{figure}

Our results also suggest that neither the presence of a bar seems more likely to have affected the integrated star formation activity of our galaxies. In Fig.~\ref{fig_bar}, we compare the distributions of sSFRs of the barred LTGs in our sample against those without a bar of similar stellar mass (10.5$\le$log(M$_{star}$/M$\odot$)$<$11.3). It is expected that the presence of a bar would result in the temporal enhancement of the star formation activity in the central region, followed by a drop of SFR due to gas consumption and feedback. \citet{catalan17} studied the spatially resolved SFRs of the CALIFA galaxies and observed an small enhancement in the central ones in barred galaxies. Yet our results suggest that no significant differences exist in their {\it integrated} star formation activity, also according to the KS-test (P$_{KS}$=0.28). This similarity holds even if we distinguish between strong and weak bars (P$_{KS}>$0.30).  The results are also confirmed by comparing the corresponding distributions of star formation efficiency where we find no difference (P$_{KS}>$0.50; see inset panel of Fig.~\ref{fig_bar}). Those findings agree with \citet{Kim17} who studied a sample of 10,000 nearby disk galaxies and showed that although strongly barred galaxies can present some small decrease of integrated star formation \citep[also see][]{Jogee05}, the overall picture is that there is no statistical difference in the stellar populations of barred and non-barred galaxies.  A similar conclusion is reached by \citet{Cervantes17} who finds almost no difference in the gas consumption timescale of barred and unbarred galaxies, in agreement with our results regarding the star formation efficiency.

\begin{figure}
\begin{center}
\includegraphics[scale=0.47]{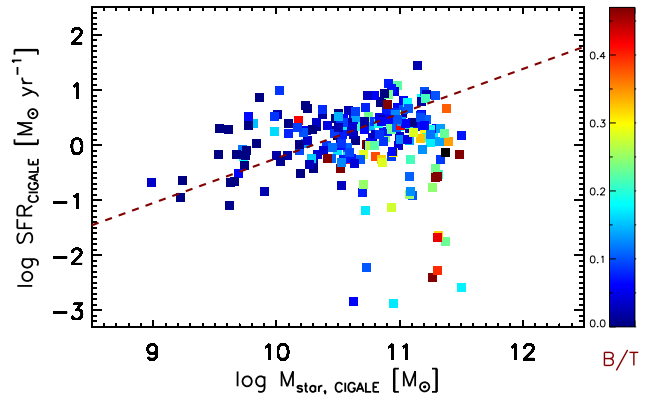}
\caption{The SFRs of 224 CALIFA late-type galaxies estimated by SED modeling versus their stellar mass, as a function of bulge-to-total luminosity rations (B/T). (A coloured version of this figure is available in the online journal)}
\label{fig_bt}
\end{center}
\end{figure}

Another important factor that has been associated with star formation quenching is the build-up of  stellar mass and as a consequence the morphological evolution with the subsequent changes in the internal galaxy dynamics. \citet{kennicutt89} showed that bulge-dominated disk galaxies are less efficiently forming stars than disk-dominated ones. Later on, \citet{Martig09} suggested the term ``morphological quenching'' to describe this process, and suggested that the presence of a spheroid (i.e. a bulge) can stabilize permanently the disk gas against gravitational collapse with result to eventually quench the star formation activity. More recently, \citet{Morishita15} studied massive galaxies at z$<$0.3 and suggested that quenching is associated with their bulge mass. As we already showed in Fig.~\ref{fig_morph}, many LTGs below the SFMS have indeed morphologies consistent with those of early-type spirals, systems that are usually dominated by large bulges. An additional way to probe the significance of bulges in the dynamics of those galaxies comes from the two-dimensional multi-component photometric decomposition of the CALIFA galaxies \citep{Mendez17}. In Fig.~\ref{fig_bt} we present again the SFR/M$_{star}$ diagram of the LTGs in our sample as a function of their bulge-to-total luminosity (B/T) ratios. LTGs below the SFMS are among those having the highest B/T's \citep[e.g. early-type spirals have B/T$>$0.2;][]{Mendez17}. 


To further ensure our previous conclusions, we are also using a machine learning classification algorithm (i.e. the {\sevensize scikit-learn LogisticRegression} as well as the Pearson correlation in {\sevensize Python}) on the LTGs in our sample, to quantify the correlation between the various galaxy parameters and quenching. To train the algorithms we introduce a binary quenching parameter (0: false and 1: true) depending on the position of each galaxy on the SFMS diagram (SF or quiescent as described in \S4.1), AGN activity (0: false, 1: true; classified either by {\sevensize CIGALE} or the emission line ratios), galaxy morphology (0: Sa-SBa, 1: Sb-SBb, and 2: Sc-SBc or above), bar (0: false, 1: weak, and 2: strong), galaxy mass (split in 5 equal mass bins with ascending order; 0-4), and finally the gas fraction (split in 6 equal mass bins with ascending order; 0-5). The results are presented in Table~\ref{tab3}, where we can see a strong anti-correlation (negative sign) between quenching and galaxy morphology and gas fraction, as well as a strong correlation (positive sign) with galaxy mass (indirectly associated with the morphology). On the other hand, both bars and AGNs display a weak correlation with the quenching of star formation, exactly as it was suggested earlier, the weakest being with the presence of a bar. It is worth mentioning here that the above results do not change with different binning in mass and gas fraction.

\begin{center}
\begin{table}
\begin{minipage}{120mm}
\caption{Correlation values of galaxy parameters with quenching}
\label{tab3}
\begin{tabular}{lcc}
\hline 
 Galaxy & Logistic & Pearson\\
 parameter & Regression & correlation\\
\hline
Mass & 1.13 & 0.32 \\
Bar & -0.28 & -0.13 \\
AGN & -0.37 & 0.04 \\
Gas fraction & -1.10 & -0.53 \\
Morphology & -1.33 & -0.45 \\
\hline
\end{tabular}
\end{minipage}
\end{table}%
\end{center}

\begin{figure*}
\begin{center}
\includegraphics[scale=0.83]{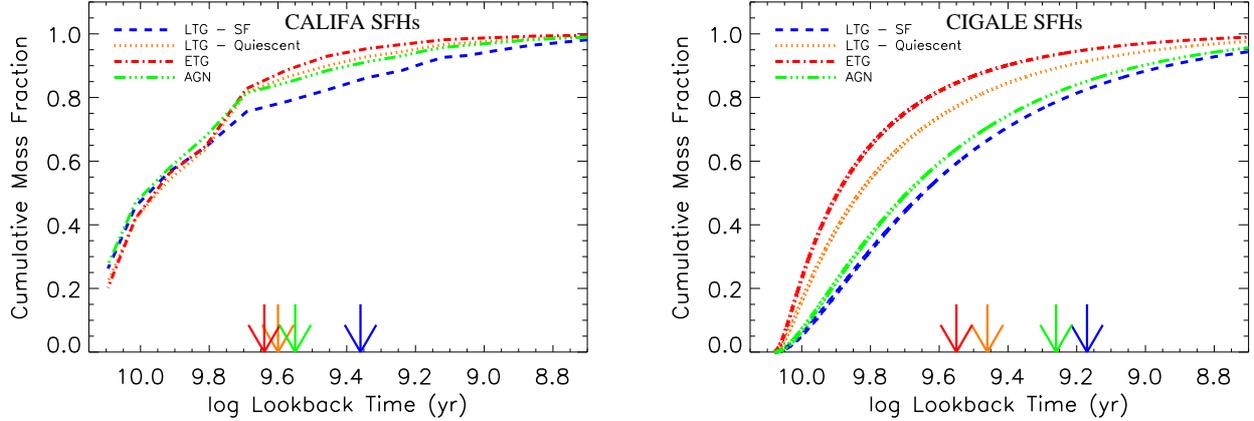}
\caption{The normalized cumulative mass distributions of the galaxies in our sample derived from the CALIFA IFU data (left panel) and the SED modelling (right panel). Star forming LTGs are shown in dashed blue lines, quiescent LTGs in dotted orange lines, ETGs in dashed dotted red lines, and AGN hosts with dashed triple dotted green lines. The vertical arrows indicate the age at which each galaxy group have built 85 per cent of their mass. Typical  normalized cumulative mass uncertainties are 0.11 and 0.16 for the left and right panels, respectively. They correspond to 0.2-0.3 Gyr uncertainties in the ages indicated by the arrows. (A coloured version of this figure is available in the online journal)}
\label{fig_sfh}
\end{center}
\end{figure*}

Finally, we examine the local environment as another mechanism to influence (enhance or reduce) the star formation activity of our galaxies. As we mentioned earlier, galaxy interactions can enhance the star formation activity of the galaxies involved by creating instabilities, and thus fragment the gas clouds, and/or enhance their gas content via accretion and merging. On the other hand, frequent and numerous interactions and/or ram pressure can have the opposite effects; such processes may strip significant amounts of interstellar gas out of the galaxies' main bodies, or even introduce shocks that may temporarily shut-down star formation. In our sample, only 2 quiescent LTGs and 3 ETGs found below the SFMS display signs of merging. On the other hand, most of the 23 galaxies located on-or-above the SFMS do not. In a comprehensive study,  \citet{catalan17} showed that there is not change in the star formation activity of the CALIFA galaxies as a function of the local galaxy density. Moreover, \citet{Barrera-Ballesteros15} analyzed the effects of galaxy merging and interactions in the star formation activity and although they found small differences in the central values, they showed that the global parameters do not change. The above results suggest that the local galaxy environment is not associated with the quenching of star formation in the CALIFA sample, having however in mind that only a small fraction of them reside in harsh environments \citep[$<$10 per cent;][]{walcher14}.

\subsection{The star formation histories}
The above results identify the mass/morphology changes and the depletion of gas, due to past star formation, as the key processes for quenching the star formation in galaxies. Other effects, such as galaxy interactions, AGN activity etc. may temporarily suppress (or enhance) the star formation activity but their effects are ephemeral. Whatever those effects are, though, they should have left their signatures on the SFHs of the galaxies involved. Such information can be drawn from the derived SFHs of our galaxies.

In both panels of Fig.~\ref{fig_sfh}, we present the normalized cumulative mass distributions for the galaxies in our sample. Typical normalized cumulative mass uncertainties are 0.11 and 0.16 for the two panels, respectively. We group our galaxies in the following sub-samples: SF late-type (marked with LTG-SF in the plot), quiescent late-type (LTG-Quiescent), early-type galaxies (ETG), and AGN hosts (AGN). To ensure the robustness of our conclusions we plot both SFHs derived by the IFU data, which are more challenging to re-construct especially for the oldest epochs (left panel; see \S3.1), and the SED models (right panel; see \S3.2). As we can see in both panels, the age at which the different galaxy groups have built most of their mass spans up to 3 Gyr. More specifically, the SF LTGs are building their stellar masses at a slower pace, whereas the ETGs have already built most of their masses ($>$85 per cent) at log(lookback/yr)$>$9.6$\pm$0.3, which corresponds to a $z$ of $>$0.5. Similarly, the quiescent LTGs have also built the largest fraction of their mass, which is contained in their bulges as opposed to SF galaxies, e.g. Bonfini et al. (in prep.), at high redshift and their mass profiles have flattened-out approximately 1.5-2 Gyr ago. The results are in agreement with \citet{GarciaBenito17} who showed that the SFHs connect with the galaxy Hubble type, with ETG masses assembled faster and at earlier cosmic times \citep[also see][]{Thomas10,McDermid15,Pacifici16}. Finally, the AGN hosts are located between the SF and quiescent LTGs, since the physics of the accretion process onto the super massive black hole does not appear to influence the evolution of the hosts and they simply comprise by $\sim$70 per cent SF and $\sim$30 per cent quiescent LTGs. As one can see the results of the SFHs of our galaxies are consistent with our earlier findings (\S4.2). Indeed, both gas consumption and the build-up of a bulge are slow processes, and have left their marks on the SFHs of the quiescent LTGs and ETGs at earlier cosmic times. \citet{LopezFernandez18} have also showed that the star formation activity of ETGs declined faster to that of LTGs. The above results are also consistent with those from \citet{eperez13} and \citet{ibarra16}. 

\section{Discussion: What quenches star formation in the CALIFA galaxies?}
Throughout this work we have listed several mechanisms that can be hold responsible to permanently ceasing (i.e. quenching) the star formation activity of galaxies. Using the powerful combination of SED modelling, IFU data, and machine learning, we attempted to disentangle and explore all the potential processes. Our results suggested that quenching is coming as a combination of gas deficiency -- most probably due to past star formation -- with the morphological evolution of the involved galaxies. The former factor favors quiescence simply because it exhausts the galaxies gas reservoir while the latter because the re-arrangement of the galaxy potential prevents the remaining gas from further collapse and hence hinders star formation. A similar conclusion has been drawn by \citet{sanchez18} for a sample of $\sim$2000 -- including 98 AGN-hosting -- {\it MaNGA} galaxies; they showed that their star formation is suppressed in an inside-out fashion by a combination of gas deficit and inefficiency. Yet this inefficiency is associated with the enhancement of the central bulge of the galaxies. Indeed, \citet{GonzalezDelgado16} showed that the sSFRs of the CALIFA galaxies increase radially outwards as galaxies quench inside-out, with the process being faster in bulge dominated systems. On the other hand, \citet{Schreiber16} examined the flattening of the SFMS at higher stellar mass star forming galaxies, and suggested that it does not depend on the existence/size of a central bulge but rather on the enhancement of the stellar mass itself. 

To obtain a more complete picture, we have also explored other important processes which have been suggested to suppress star formation and we have examined their role in quenching. For instance, the presence of an AGN and the action of bars do not seem to have strong effects on the SFHs of the involved galaxies, although temporarily they may influence their star formation activity. Indeed, Ibarra-Medel et al. (in prep.) are using a new method to study the fossil records (i.e. they use stellar population synthesis models to stress changes in the galaxy properties, such as the SFR, throughout the SFH of each galaxy) of the {\it MaNGA} galaxies, and find that AGN-hosts are {\it oscillating} between the SFMS and the green-valley, as the AGN re-ignites each time. On the other hand, there might also be a non-trivial connection between the triggering of the AGN activity in galaxies with high stellar masses, occurring at the same time with the quenching of their star formation; the growth of the central bulge, which we consider responsible for inhibiting star formation, could have been causing the triggering of AGN as testified by the bulge-SMBH growth connection \citep[e.g. the M$_{BH}-\sigma$ relation;][]{Merritt01}. Finally, we have found no correlation between the galaxy environment (or galaxy mergers) and the quenching of star formation in our galaxies. Yet, we have to stress here that the CALIFA survey originally did not include galaxies in extreme environments, e.g. galaxy groups and near the centers of clusters.

\section{Conclusions}
In this article we are presenting the integrated properties of 835 galaxies of the CALIFA survey. Those properties were derived using the available optical observations and state-of-the-art theoretical modeling. We perform comparisons between our model-derived galaxy properties and those from the integral field unit data, and analyzed the results to conclude the following:

\begin{itemize}
\item The star formation rates of quiescent galaxies, based on their spectrum derived H$\alpha$ luminosities and the stellar population synthesis models suffer from heavy contamination from evolved stars. Thus, future studies aiming to use the SFR of the galaxies in the survey should employ those estimated from SED modeling.

\item Galaxy quenching results from a combination of gas deficiency and the inefficiency of the remaining gas to form stars. After a comprehensive analysis, we attribute the latter to morphological evolution (i.e. the build-up of the bulge component).

\item Exploring other potential suppression mechanisms, such as AGN activity, the action of bars, galaxy mergers, and the local galaxy environment, we find no strong evidence that could associate them with permanent quenching of star formation.

\item The study of the star formation histories of our galaxies as derived from SED fitting and stellar population synthesis confirms the above findings.
\end{itemize}
\section*{Acknowledgements}

TB would like to acknowledge support from the CONACyT Research Fellowships program. T.D.-S. acknowledges support from ALMA-CONICYT project 31130005 and FONDECYT regular project 1151239. TB would like to thank G. Zachou for the support while writing this article. BCS acknowledges financial support through PAPIIT project IA103517 from DGAPA-UNAM. SFS thanks the support by CONACYT grant CB-285080 and PAPIIT-DGAPA-IA101217 (UNAM) project. The authors would like to thank the referee for the comments that helped improve this article. This study  uses data provided by the Calar-Alto Legacy Integral Field Area (CALIFA) survey (http://califa.caha.es/). CALIFA is the first legacy survey performed at Calar Alto. The CALIFA collaboration would like to thank the IAA-CSIC and MPIA-MPG as major partners of the observatory, and CAHA itself, for the unique access to telescope time and support in manpower and infrastructures. The CALIFA collaboration also thanks the CAHA staff for the dedication to this project. Based on observations collected at the Centro Astron\'omico Hispano Alem\'an (CAHA) at Calar Alto, operated jointly by the Max-Planck-Institut f\"ur Astronomie and the Instituto de Astrof\'\i sica de Andaluc\'\i a  (CSIC).




\bibliographystyle{mnras}
\bibliography{CALIFAI.bib} 

\begin{thebibliography}{}
\makeatletter
\relax
\def\mn@urlcharsother{\let\do\@makeother \do\$\do\&\do\#\do\^\do\_\do\%\do\~}
\def\mn@doi{\begingroup\mn@urlcharsother \@ifnextchar [ {\mn@doi@}
  {\mn@doi@[]}}
\def\mn@doi@[#1]#2{\def\@tempa{#1}\ifx\@tempa\@empty \href
  {http://dx.doi.org/#2} {doi:#2}\else \href {http://dx.doi.org/#2} {#1}\fi
  \endgroup}
\def\mn@eprint#1#2{\mn@eprint@#1:#2::\@nil}
\def\mn@eprint@arXiv#1{\href {http://arxiv.org/abs/#1} {{\tt arXiv:#1}}}
\def\mn@eprint@dblp#1{\href {http://dblp.uni-trier.de/rec/bibtex/#1.xml}
  {dblp:#1}}
\def\mn@eprint@#1:#2:#3:#4\@nil{\def\@tempa {#1}\def\@tempb {#2}\def\@tempc
  {#3}\ifx \@tempc \@empty \let \@tempc \@tempb \let \@tempb \@tempa \fi \ifx
  \@tempb \@empty \def\@tempb {arXiv}\fi \@ifundefined
  {mn@eprint@\@tempb}{\@tempb:\@tempc}{\expandafter \expandafter \csname
  mn@eprint@\@tempb\endcsname \expandafter{\@tempc}}}

\bibitem[\protect\citeauthoryear{{Abazajian} et~al.,}{{Abazajian}
  et~al.}{2009}]{abazajian09}
{Abazajian} K.~N.,  et~al., 2009, \mn@doi [\apjs]
  {10.1088/0067-0049/182/2/543}, \href
  {http://adsabs.harvard.edu/abs/2009ApJS..182..543A} {182, 543}

\bibitem[\protect\citeauthoryear{{Alatalo} et~al.,}{{Alatalo}
  et~al.}{2015}]{Alatalo15}
{Alatalo} K.,  et~al., 2015, \mn@doi [\apj] {10.1088/0004-637X/798/1/31}, \href
  {http://adsabs.harvard.edu/abs/2015ApJ...798...31A} {798, 31}

\bibitem[\protect\citeauthoryear{{Alatalo} et~al.,}{{Alatalo}
  et~al.}{2016}]{Alatalo16}
{Alatalo} K.,  et~al., 2016, \mn@doi [\apjs] {10.3847/0067-0049/224/2/38},
  \href {http://adsabs.harvard.edu/abs/2016ApJS..224...38A} {224, 38}

\bibitem[\protect\citeauthoryear{{Barrera-Ballesteros}
  et~al.,}{{Barrera-Ballesteros} et~al.}{2015}]{Barrera-Ballesteros15}
{Barrera-Ballesteros} J.~K.,  et~al., 2015, \mn@doi [\aap]
  {10.1051/0004-6361/201425397}, \href
  {http://adsabs.harvard.edu/abs/2015A%26A...579A..45B} {579, A45}

\bibitem[\protect\citeauthoryear{{Barrera-Ballesteros}
  et~al.,}{{Barrera-Ballesteros} et~al.}{2018}]{Barrera-Ballesteros18}
{Barrera-Ballesteros} J.~K.,  et~al., 2018, \mn@doi [\apj]
  {10.3847/1538-4357/aa9b31}, \href
  {http://adsabs.harvard.edu/abs/2018ApJ...852...74B} {852, 74}

\bibitem[\protect\citeauthoryear{{Binette}, {Magris}, {Stasi{\'n}ska}  \&
  {Bruzual}}{{Binette} et~al.}{1994}]{Binette94}
{Binette} L.,  {Magris} C.~G.,  {Stasi{\'n}ska} G.,   {Bruzual} A.~G.,  1994,
  \aap, \href {http://adsabs.harvard.edu/abs/1994A%26A...292...13B} {292, 13}

\bibitem[\protect\citeauthoryear{{Bitsakis}, {Charmandaris}, {da Cunha},
  {D{\'{\i}}az-Santos}, {Le Floc'h}  \& {Magdis}}{{Bitsakis}
  et~al.}{2011}]{Bitsakis11}
{Bitsakis} T.,  {Charmandaris} V.,  {da Cunha} E.,  {D{\'{\i}}az-Santos} T.,
  {Le Floc'h} E.,   {Magdis} G.,  2011, \mn@doi [\aap]
  {10.1051/0004-6361/201117355}, \href
  {http://adsabs.harvard.edu/abs/2011A%26A...533A.142B} {533, A142}

\bibitem[\protect\citeauthoryear{{Bitsakis}, {Dultzin}, {Ciesla}, {Krongold},
  {Charmandaris}  \& {Zezas}}{{Bitsakis} et~al.}{2015}]{Bitsakis15}
{Bitsakis} T.,  {Dultzin} D.,  {Ciesla} L.,  {Krongold} Y.,  {Charmandaris} V.,
    {Zezas} A.,  2015, \mn@doi [\mnras] {10.1093/mnras/stv755}, \href
  {http://adsabs.harvard.edu/abs/2015MNRAS.450.3114B} {450, 3114}

\bibitem[\protect\citeauthoryear{{Bitsakis} et~al.,}{{Bitsakis}
  et~al.}{2016}]{Bitsakis16}
{Bitsakis} T.,  et~al., 2016, \mn@doi [\mnras] {10.1093/mnras/stw686}, \href
  {http://adsabs.harvard.edu/abs/2016MNRAS.459..957B} {459, 957}

\bibitem[\protect\citeauthoryear{{Bolatto} et~al.,}{{Bolatto}
  et~al.}{2017}]{Bolatto17}
{Bolatto} A.~D.,  et~al., 2017, \mn@doi [\apj] {10.3847/1538-4357/aa86aa},
  \href {http://adsabs.harvard.edu/abs/2017ApJ...846..159B} {846, 159}

\bibitem[\protect\citeauthoryear{{Bundy} et~al.,}{{Bundy}
  et~al.}{2015}]{Bundy15}
{Bundy} K.,  et~al., 2015, \mn@doi [\apj] {10.1088/0004-637X/798/1/7}, \href
  {http://adsabs.harvard.edu/abs/2015ApJ...798....7B} {798, 7}

\bibitem[\protect\citeauthoryear{{Cano-D{\'{\i}}az} et~al.,}{{Cano-D{\'{\i}}az}
  et~al.}{2016}]{mariana16}
{Cano-D{\'{\i}}az} M.,  et~al., 2016, \mn@doi [\apjl]
  {10.3847/2041-8205/821/2/L26}, \href
  {http://adsabs.harvard.edu/abs/2016ApJ...821L..26C} {821, L26}

\bibitem[\protect\citeauthoryear{{Catal{\'a}n-Torrecilla}
  et~al.,}{{Catal{\'a}n-Torrecilla} et~al.}{2015}]{catalan15}
{Catal{\'a}n-Torrecilla} C.,  et~al., 2015, \mn@doi [\aap]
  {10.1051/0004-6361/201526023}, \href
  {http://adsabs.harvard.edu/abs/2015A%26A...584A..87C} {584, A87}

\bibitem[\protect\citeauthoryear{{Catal{\'a}n-Torrecilla}
  et~al.,}{{Catal{\'a}n-Torrecilla} et~al.}{2017}]{catalan17}
{Catal{\'a}n-Torrecilla} C.,  et~al., 2017, \mn@doi [\apj]
  {10.3847/1538-4357/aa8a6d}, \href
  {http://adsabs.harvard.edu/abs/2017ApJ...848...87C} {848, 87}

\bibitem[\protect\citeauthoryear{{Cervantes Sodi}}{{Cervantes
  Sodi}}{2017}]{Cervantes17}
{Cervantes Sodi} B.,  2017, \mn@doi [\apj] {10.3847/1538-4357/835/1/80}, \href
  {http://adsabs.harvard.edu/abs/2017ApJ...835...80C} {835, 80}

\bibitem[\protect\citeauthoryear{{Cid Fernandes} et~al.,}{{Cid Fernandes}
  et~al.}{2013}]{cid-fernandes13}
{Cid Fernandes} R.,  et~al., 2013, \mn@doi [\aap]
  {10.1051/0004-6361/201220616}, \href
  {http://adsabs.harvard.edu/abs/2013A%26A...557A..86C} {557, A86}

\bibitem[\protect\citeauthoryear{{Ciesla} et~al.,}{{Ciesla}
  et~al.}{2015}]{Ciesla15}
{Ciesla} L.,  et~al., 2015, \mn@doi [\aap] {10.1051/0004-6361/201425252}, \href
  {http://adsabs.harvard.edu/abs/2015A%26A...576A..10C} {576, A10}

\bibitem[\protect\citeauthoryear{{Ciesla} et~al.,}{{Ciesla}
  et~al.}{2016}]{ciesla16}
{Ciesla} L.,  et~al., 2016, \mn@doi [\aap] {10.1051/0004-6361/201527107}, \href
  {http://adsabs.harvard.edu/abs/2016A%26A...585A..43C} {585, A43}

\bibitem[\protect\citeauthoryear{{Ciesla}, {Elbaz}  \& {Fensch}}{{Ciesla}
  et~al.}{2017}]{ciesla17}
{Ciesla} L.,  {Elbaz} D.,   {Fensch} J.,  2017, \mn@doi [\aap]
  {10.1051/0004-6361/201731036}, \href
  {http://adsabs.harvard.edu/abs/2017A%26A...608A..41C} {608, A41}

\bibitem[\protect\citeauthoryear{{Di Matteo}, {Springel}  \& {Hernquist}}{{Di
  Matteo} et~al.}{2005}]{dimatteo05}
{Di Matteo} T.,  {Springel} V.,   {Hernquist} L.,  2005, \mn@doi [\nat]
  {10.1038/nature03335}, \href
  {http://adsabs.harvard.edu/abs/2005Natur.433..604D} {433, 604}

\bibitem[\protect\citeauthoryear{{Elbaz} et~al.,}{{Elbaz}
  et~al.}{2007}]{Elbaz07}
{Elbaz} D.,  et~al., 2007, \mn@doi [\aap] {10.1051/0004-6361:20077525}, \href
  {http://adsabs.harvard.edu/abs/2007A%26A...468...33E} {468, 33}

\bibitem[\protect\citeauthoryear{{Ellison}, {S{\'a}nchez}, {Ibarra-Medel},
  {Antonio}, {Mendel}  \& {Barrera-Ballesteros}}{{Ellison}
  et~al.}{2018}]{Ellison18}
{Ellison} S.~L.,  {S{\'a}nchez} S.~F.,  {Ibarra-Medel} H.,  {Antonio} B.,
  {Mendel} J.~T.,   {Barrera-Ballesteros} J.,  2018, \mn@doi [\mnras]
  {10.1093/mnras/stx2882}, \href
  {http://adsabs.harvard.edu/abs/2018MNRAS.474.2039E} {474, 2039}

\bibitem[\protect\citeauthoryear{{Fabian}}{{Fabian}}{2012}]{Fabian12}
{Fabian} A.~C.,  2012, \mn@doi [\araa] {10.1146/annurev-astro-081811-125521},
  \href {http://adsabs.harvard.edu/abs/2012ARA%26A..50..455F} {50, 455}

\bibitem[\protect\citeauthoryear{{Galbany} et~al.,}{{Galbany}
  et~al.}{2017}]{Galbany17}
{Galbany} L.,  et~al., 2017, \mn@doi [\mnras] {10.1093/mnras/stx367}, \href
  {http://adsabs.harvard.edu/abs/2017MNRAS.468..628G} {468, 628}

\bibitem[\protect\citeauthoryear{{Galbany} et~al.,}{{Galbany}
  et~al.}{2018}]{Galbany18}
{Galbany} L.,  et~al., 2018, \mn@doi [\apj] {10.3847/1538-4357/aaaf20}, \href
  {http://adsabs.harvard.edu/abs/2018ApJ...855..107G} {855, 107}

\bibitem[\protect\citeauthoryear{{Garc{\'{\i}}a-Benito}
  et~al.,}{{Garc{\'{\i}}a-Benito} et~al.}{2017}]{GarciaBenito17}
{Garc{\'{\i}}a-Benito} R.,  et~al., 2017, \mn@doi [\aap]
  {10.1051/0004-6361/201731357}, \href
  {http://adsabs.harvard.edu/abs/2017A%26A...608A..27G} {608, A27}

\bibitem[\protect\citeauthoryear{{Gomes} et~al.,}{{Gomes}
  et~al.}{2016}]{Gomes16}
{Gomes} J.~M.,  et~al., 2016, \mn@doi [\aap] {10.1051/0004-6361/201525976},
  \href {http://adsabs.harvard.edu/abs/2016A%26A...588A..68G} {588, A68}

\bibitem[\protect\citeauthoryear{{Gonz{\'a}lez Delgado} et~al.,}{{Gonz{\'a}lez
  Delgado} et~al.}{2016a}]{rosa16a}
{Gonz{\'a}lez Delgado} R.~M.,  et~al., 2016a, \mn@doi [\aap]
  {10.1051/0004-6361/201628174}, \href
  {http://adsabs.harvard.edu/abs/2016A%26A...590A..44G} {590, A44}

\bibitem[\protect\citeauthoryear{{Gonz{\'a}lez Delgado} et~al.,}{{Gonz{\'a}lez
  Delgado} et~al.}{2016b}]{GonzalezDelgado16}
{Gonz{\'a}lez Delgado} R.~M.,  et~al., 2016b, \mn@doi [\aap]
  {10.1051/0004-6361/201628174}, \href
  {http://adsabs.harvard.edu/abs/2016A%26A...590A..44G} {590, A44}

\bibitem[\protect\citeauthoryear{{Hsieh} et~al.,}{{Hsieh}
  et~al.}{2017}]{Hsieh17}
{Hsieh} B.~C.,  et~al., 2017, \mn@doi [\apjl] {10.3847/2041-8213/aa9d80}, \href
  {http://adsabs.harvard.edu/abs/2017ApJ...851L..24H} {851, L24}

\bibitem[\protect\citeauthoryear{{Ibarra-Medel} et~al.,}{{Ibarra-Medel}
  et~al.}{2016}]{ibarra16}
{Ibarra-Medel} H.~J.,  et~al., 2016, \mn@doi [\mnras] {10.1093/mnras/stw2126},
  \href {http://adsabs.harvard.edu/abs/2016MNRAS.463.2799I} {463, 2799}

\bibitem[\protect\citeauthoryear{{Jogee}, {Scoville}  \& {Kenney}}{{Jogee}
  et~al.}{2005}]{Jogee05}
{Jogee} S.,  {Scoville} N.,   {Kenney} J.~D.~P.,  2005, \mn@doi [\apj]
  {10.1086/432106}, \href {http://adsabs.harvard.edu/abs/2005ApJ...630..837J}
  {630, 837}

\bibitem[\protect\citeauthoryear{{Kehrig} et~al.,}{{Kehrig}
  et~al.}{2012}]{Kehrig12}
{Kehrig} C.,  et~al., 2012, \mn@doi [\aap] {10.1051/0004-6361/201118357}, \href
  {http://adsabs.harvard.edu/abs/2012A%26A...540A..11K} {540, A11}

\bibitem[\protect\citeauthoryear{{Kennicutt}}{{Kennicutt}}{1989}]{kennicutt89}
{Kennicutt} Jr. R.~C.,  1989, \mn@doi [\apj] {10.1086/167834}, \href
  {http://adsabs.harvard.edu/abs/1989ApJ...344..685K} {344, 685}

\bibitem[\protect\citeauthoryear{{Kewley}, {Groves}, {Kauffmann}  \&
  {Heckman}}{{Kewley} et~al.}{2006}]{kewley06}
{Kewley} L.~J.,  {Groves} B.,  {Kauffmann} G.,   {Heckman} T.,  2006, \mn@doi
  [\mnras] {10.1111/j.1365-2966.2006.10859.x}, \href
  {http://adsabs.harvard.edu/abs/2006MNRAS.372..961K} {372, 961}

\bibitem[\protect\citeauthoryear{{Kim}, {Hwang}, {Chung}, {Lee}, {Park},
  {Cervantes Sodi}  \& {Kim}}{{Kim} et~al.}{2017}]{Kim17}
{Kim} E.,  {Hwang} H.~S.,  {Chung} H.,  {Lee} G.-H.,  {Park} C.,  {Cervantes
  Sodi} B.,   {Kim} S.~S.,  2017, \mn@doi [\apj] {10.3847/1538-4357/aa80db},
  \href {http://adsabs.harvard.edu/abs/2017ApJ...845...93K} {845, 93}

\bibitem[\protect\citeauthoryear{{Lacerda} et~al.,}{{Lacerda}
  et~al.}{2018}]{Lacerda18}
{Lacerda} E.~A.~D.,  et~al., 2018, \mn@doi [\mnras] {10.1093/mnras/stx3022},
  \href {http://adsabs.harvard.edu/abs/2018MNRAS.474.3727L} {474, 3727}

\bibitem[\protect\citeauthoryear{{L{\'o}pez Fern{\'a}ndez} et~al.,}{{L{\'o}pez
  Fern{\'a}ndez} et~al.}{2016}]{LopezFernandez16}
{L{\'o}pez Fern{\'a}ndez} R.,  et~al., 2016, \mn@doi [\mnras]
  {10.1093/mnras/stw260}, \href
  {http://adsabs.harvard.edu/abs/2016MNRAS.458..184L} {458, 184}

\bibitem[\protect\citeauthoryear{{L{\'o}pez Fern{\'a}ndez} et~al.,}{{L{\'o}pez
  Fern{\'a}ndez} et~al.}{2018}]{LopezFernandez18}
{L{\'o}pez Fern{\'a}ndez} R.,  et~al., 2018, preprint, \href
  {http://adsabs.harvard.edu/abs/2018arXiv180210118L} {} (\mn@eprint {arXiv}
  {1802.10118})

\bibitem[\protect\citeauthoryear{{Martig}, {Bournaud}, {Teyssier}  \&
  {Dekel}}{{Martig} et~al.}{2009}]{Martig09}
{Martig} M.,  {Bournaud} F.,  {Teyssier} R.,   {Dekel} A.,  2009, \mn@doi
  [\apj] {10.1088/0004-637X/707/1/250}, \href
  {http://adsabs.harvard.edu/abs/2009ApJ...707..250M} {707, 250}

\bibitem[\protect\citeauthoryear{{McDermid} et~al.,}{{McDermid}
  et~al.}{2015}]{McDermid15}
{McDermid} R.~M.,  et~al., 2015, \mn@doi [\mnras] {10.1093/mnras/stv105}, \href
  {http://adsabs.harvard.edu/abs/2015MNRAS.448.3484M} {448, 3484}

\bibitem[\protect\citeauthoryear{{M{\'e}ndez-Abreu} et~al.,}{{M{\'e}ndez-Abreu}
  et~al.}{2017}]{Mendez17}
{M{\'e}ndez-Abreu} J.,  et~al., 2017, \mn@doi [\aap]
  {10.1051/0004-6361/201629525}, \href
  {http://adsabs.harvard.edu/abs/2017A%26A...598A..32M} {598, A32}

\bibitem[\protect\citeauthoryear{{Merritt} \& {Ferrarese}}{{Merritt} \&
  {Ferrarese}}{2001}]{Merritt01}
{Merritt} D.,  {Ferrarese} L.,  2001, \mn@doi [\apj] {10.1086/318372}, \href
  {http://adsabs.harvard.edu/abs/2001ApJ...547..140M} {547, 140}

\bibitem[\protect\citeauthoryear{{Morishita}, {Ichikawa}, {Noguchi}, {Akiyama},
  {Patel}, {Kajisawa}  \& {Obata}}{{Morishita} et~al.}{2015}]{Morishita15}
{Morishita} T.,  {Ichikawa} T.,  {Noguchi} M.,  {Akiyama} M.,  {Patel} S.~G.,
  {Kajisawa} M.,   {Obata} T.,  2015, \mn@doi [\apj]
  {10.1088/0004-637X/805/1/34}, \href
  {http://adsabs.harvard.edu/abs/2015ApJ...805...34M} {805, 34}

\bibitem[\protect\citeauthoryear{{Morrissey} \& {GALEX Science
  Team}}{{Morrissey} \& {GALEX Science Team}}{2005}]{morrissey05}
{Morrissey} P.,  {GALEX Science Team} 2005, in American Astronomical Society
  Meeting Abstracts. p.~1454

\bibitem[\protect\citeauthoryear{{Noll}, {Burgarella}, {Giovannoli}, {Buat},
  {Marcillac}  \& {Mu{\~n}oz-Mateos}}{{Noll} et~al.}{2009}]{Noll09}
{Noll} S.,  {Burgarella} D.,  {Giovannoli} E.,  {Buat} V.,  {Marcillac} D.,
  {Mu{\~n}oz-Mateos} J.~C.,  2009, \mn@doi [\aap]
  {10.1051/0004-6361/200912497}, \href
  {http://adsabs.harvard.edu/abs/2009A%26A...507.1793N} {507, 1793}

\bibitem[\protect\citeauthoryear{{Okada}, {Murakami}, {Yasuda}  et~al.}{{Okada}
  et~al.}{2008}]{okada08}
{Okada} Y.,  {Murakami} N.,  {Yasuda} A.,   et~al., 2008

\bibitem[\protect\citeauthoryear{{Pacifici} et~al.,}{{Pacifici}
  et~al.}{2016}]{Pacifici16}
{Pacifici} C.,  et~al., 2016, \mn@doi [\apj] {10.3847/0004-637X/832/1/79},
  \href {http://adsabs.harvard.edu/abs/2016ApJ...832...79P} {832, 79}

\bibitem[\protect\citeauthoryear{{Papaderos} et~al.,}{{Papaderos}
  et~al.}{2013}]{Papaderos13}
{Papaderos} P.,  et~al., 2013, \mn@doi [\aap] {10.1051/0004-6361/201321681},
  \href {http://adsabs.harvard.edu/abs/2013A%26A...555L...1P} {555, L1}

\bibitem[\protect\citeauthoryear{{Peng}, {Maiolino}  \& {Cochrane}}{{Peng}
  et~al.}{2015}]{Peng15}
{Peng} Y.,  {Maiolino} R.,   {Cochrane} R.,  2015, \mn@doi [\nat]
  {10.1038/nature14439}, \href
  {http://adsabs.harvard.edu/abs/2015Natur.521..192P} {521, 192}

\bibitem[\protect\citeauthoryear{{P{\'e}rez} et~al.,}{{P{\'e}rez}
  et~al.}{2013}]{eperez13}
{P{\'e}rez} E.,  et~al., 2013, \mn@doi [\apjl] {10.1088/2041-8205/764/1/L1},
  \href {http://adsabs.harvard.edu/abs/2013ApJ...764L...1P} {764, L1}

\bibitem[\protect\citeauthoryear{{Quilis}, {Moore}  \& {Bower}}{{Quilis}
  et~al.}{2000}]{Quillis00}
{Quilis} V.,  {Moore} B.,   {Bower} R.,  2000, \mn@doi [Science]
  {10.1126/science.288.5471.1617}, \href
  {http://adsabs.harvard.edu/abs/2000Sci...288.1617Q} {288, 1617}

\bibitem[\protect\citeauthoryear{{Saintonge} et~al.,}{{Saintonge}
  et~al.}{2016}]{Saintonge16}
{Saintonge} A.,  et~al., 2016, \mn@doi [\mnras] {10.1093/mnras/stw1715}, \href
  {http://adsabs.harvard.edu/abs/2016MNRAS.462.1749S} {462, 1749}

\bibitem[\protect\citeauthoryear{{S{\'a}nchez} et~al.,}{{S{\'a}nchez}
  et~al.}{2012}]{sanchez12a}
{S{\'a}nchez} S.~F.,  et~al., 2012, \mn@doi [\aap]
  {10.1051/0004-6361/201117353}, \href
  {http://adsabs.harvard.edu/abs/2012A%26A...538A...8S} {538, A8}

\bibitem[\protect\citeauthoryear{{S{\'a}nchez} et~al.,}{{S{\'a}nchez}
  et~al.}{2016a}]{Sanchez16a}
{S{\'a}nchez} S.~F.,  et~al., 2016a, \rmxaa, \href
  {http://adsabs.harvard.edu/abs/2016RMxAA..52...21S} {52, 21}

\bibitem[\protect\citeauthoryear{{S{\'a}nchez} et~al.,}{{S{\'a}nchez}
  et~al.}{2016b}]{sanchez16b}
{S{\'a}nchez} S.~F.,  et~al., 2016b, \rmxaa, \href
  {http://adsabs.harvard.edu/abs/2016RMxAA..52..171S} {52, 171}

\bibitem[\protect\citeauthoryear{{S{\'a}nchez} et~al.,}{{S{\'a}nchez}
  et~al.}{2016c}]{sanchez16c}
{S{\'a}nchez} S.~F.,  et~al., 2016c, \mn@doi [\aap]
  {10.1051/0004-6361/201628661}, \href
  {http://adsabs.harvard.edu/abs/2016A%26A...594A..36S} {594, A36}

\bibitem[\protect\citeauthoryear{{S{\'a}nchez} et~al.,}{{S{\'a}nchez}
  et~al.}{2018}]{sanchez18}
{S{\'a}nchez} S.~F.,  et~al., 2018, \rmxaa, \href
  {http://adsabs.harvard.edu/abs/2018RMxAA..54..217S} {54, 217}

\bibitem[\protect\citeauthoryear{{Sarzi} et~al.,}{{Sarzi}
  et~al.}{2010}]{Sarzi10}
{Sarzi} M.,  et~al., 2010, \mn@doi [\mnras] {10.1111/j.1365-2966.2009.16039.x},
  \href {http://adsabs.harvard.edu/abs/2010MNRAS.402.2187S} {402, 2187}

\bibitem[\protect\citeauthoryear{{Schreiber}, {Elbaz}, {Pannella}, {Ciesla},
  {Wang}, {Koekemoer}, {Rafelski}  \& {Daddi}}{{Schreiber}
  et~al.}{2016}]{Schreiber16}
{Schreiber} C.,  {Elbaz} D.,  {Pannella} M.,  {Ciesla} L.,  {Wang} T.,
  {Koekemoer} A.,  {Rafelski} M.,   {Daddi} E.,  2016, \mn@doi [\aap]
  {10.1051/0004-6361/201527200}, \href
  {http://adsabs.harvard.edu/abs/2016A%26A...589A..35S} {589, A35}

\bibitem[\protect\citeauthoryear{{Scoville} et~al.,}{{Scoville}
  et~al.}{2017}]{Scoville17}
{Scoville} N.,  et~al., 2017, \mn@doi [\apj] {10.3847/1538-4357/aa61a0}, \href
  {http://adsabs.harvard.edu/abs/2017ApJ...837..150S} {837, 150}

\bibitem[\protect\citeauthoryear{{Singh} et~al.,}{{Singh}
  et~al.}{2013}]{Singh13}
{Singh} R.,  et~al., 2013, \mn@doi [\aap] {10.1051/0004-6361/201322062}, \href
  {http://adsabs.harvard.edu/abs/2013A%26A...558A..43S} {558, A43}

\bibitem[\protect\citeauthoryear{{Skrutskie} et~al.,}{{Skrutskie}
  et~al.}{2006}]{skrutskie06}
{Skrutskie} M.~F.,  et~al., 2006, \mn@doi [\aj] {10.1086/498708}, \href
  {http://adsabs.harvard.edu/abs/2006AJ....131.1163S} {131, 1163}

\bibitem[\protect\citeauthoryear{{Smith}, {Struck}, {Hancock}, {Appleton},
  {Charmandaris}  \& {Reach}}{{Smith} et~al.}{2007}]{Smith07}
{Smith} B.~J.,  {Struck} C.,  {Hancock} M.,  {Appleton} P.~N.,  {Charmandaris}
  V.,   {Reach} W.~T.,  2007, \mn@doi [\aj] {10.1086/510350}, \href
  {http://adsabs.harvard.edu/abs/2007AJ....133..791S} {133, 791}

\bibitem[\protect\citeauthoryear{{Stasi{\'n}ska}, {Vale Asari}, {Cid
  Fernandes}, {Gomes}, {Schlickmann}, {Mateus}, {Schoenell}  \&
  {Sodr{\'e}}}{{Stasi{\'n}ska} et~al.}{2008}]{stas08}
{Stasi{\'n}ska} G.,  {Vale Asari} N.,  {Cid Fernandes} R.,  {Gomes} J.~M.,
  {Schlickmann} M.,  {Mateus} A.,  {Schoenell} W.,   {Sodr{\'e}} Jr. L.,  2008,
  \mn@doi [\mnras] {10.1111/j.1745-3933.2008.00550.x}, \href
  {http://adsabs.harvard.edu/abs/2008MNRAS.391L..29S} {391, L29}

\bibitem[\protect\citeauthoryear{{Struck}}{{Struck}}{1999}]{Struck99}
{Struck} C.,  1999, \mn@doi [\physrep] {10.1016/S0370-1573(99)00030-7}, \href
  {http://adsabs.harvard.edu/abs/1999PhR...321....1S} {321, 1}

\bibitem[\protect\citeauthoryear{{Thomas}, {Maraston}, {Schawinski}, {Sarzi}
  \& {Silk}}{{Thomas} et~al.}{2010}]{Thomas10}
{Thomas} D.,  {Maraston} C.,  {Schawinski} K.,  {Sarzi} M.,   {Silk} J.,  2010,
  \mn@doi [\mnras] {10.1111/j.1365-2966.2010.16427.x}, \href
  {http://adsabs.harvard.edu/abs/2010MNRAS.404.1775T} {404, 1775}

\bibitem[\protect\citeauthoryear{{Walcher} et~al.,}{{Walcher}
  et~al.}{2014}]{walcher14}
{Walcher} C.~J.,  et~al., 2014, \mn@doi [\aap] {10.1051/0004-6361/201424198},
  \href {http://adsabs.harvard.edu/abs/2014A%26A...569A...1W} {569, A1}

\bibitem[\protect\citeauthoryear{{Wright} et~al.,}{{Wright}
  et~al.}{2010}]{wright10}
{Wright} E.~L.,  et~al., 2010, \mn@doi [\aj] {10.1088/0004-6256/140/6/1868},
  \href {http://adsabs.harvard.edu/abs/2010AJ....140.1868W} {140, 1868}

\bibitem[\protect\citeauthoryear{{Zaragoza-Cardiel}, {Smith}, {Rosado},
  {Beckman}, {Bitsakis}, {Camps-Fari{\~n}a}, {Font}  \&
  {Cox}}{{Zaragoza-Cardiel} et~al.}{2018}]{Zaragoza18}
{Zaragoza-Cardiel} J.,  {Smith} B.~J.,  {Rosado} M.,  {Beckman} J.~E.,
  {Bitsakis} T.,  {Camps-Fari{\~n}a} A.,  {Font} J.,   {Cox} I.~S.,  2018,
  \mn@doi [\apjs] {10.3847/1538-4365/aaa255}, \href
  {http://adsabs.harvard.edu/abs/2018ApJS..234...35Z} {234, 35}

\makeatother
\end{thebibliography}






\bsp	
\label{lastpage}
\end{document}